\documentstyle[prb,aps,epsf,floats,goodfloat]{revtex}

\newcommand{\bc}{\begin{center}}
\newcommand{\ec}{\end{center}}
\newcommand{\be}{\begin{equation}}
\newcommand{\ee}{\end{equation}}
\newcommand{\beqn}{\begin{eqnarray}}
\newcommand{\eeqn}{\end{eqnarray}}
\begin{document}
\draft

\twocolumn[\hsize\textwidth\columnwidth\hsize\csname@twocolumnfalse%
\endcsname

\title{
Critical Behavior and Griffiths-McCoy Singularities
in the\\ Two-Dimensional Random Quantum Ising Ferromagnet
}

\author{C. Pich and A. P. Young}
\address{Department of Physics, University of California, Santa Cruz, 
CA 95064}

\author{H. Rieger$^1$ and N. Kawashima$^{2}$\cite{lala}}
\address{$^1$NIC c/o Forschungszentrum J\"ulich, 52425 J\"ulich, Germany}
\address{$^2$Department of Physics, Toho University, Miyama 2-2-1,
Funabashi 274, Japan}

\date{\today}

\maketitle

\begin{abstract}
  We study the quantum phase transition in the two-dimensional random
  Ising model in a transverse field by Monte Carlo simulations. We
  find results similar to those known analytically in one-dimension.
  At the critical point, the dynamical exponent is infinite and the
  typical correlation function decays with a stretched exponential
  dependence on distance. Away from the critical point there are
  Griffiths-McCoy singularities, characterized by a single,
  continuously varying exponent, $z^\prime$, which diverges at the
  critical point, as in one-dimension. Consequently, the zero
  temperature susceptibility diverges for a {\em range} of parameters
  about the transition.
\end{abstract}

]


Though {\em classical} phase transitions occurring at finite
temperature are very well understood, our knowledge of {\em quantum}
transitions at $T=0$ is relatively poor, at least for systems with quenched
disorder. There is, however, considerable interest in these systems since
they (i)
exhibit new universality classes, and (ii) display
``Griffiths-McCoy''\cite{griffiths,mccoy}
singularities even away from the critical point, due to rare regions 
with stronger than average interactions.

Just as the simplest model with a
classical phase transition is the Ising model,
the simplest random model with a quantum transition is
arguably the Ising model in a transverse field whose Hamiltonian is given by
\begin{equation}
{\cal H} = -\sum_{\langle i,j\rangle} J_{ij} \sigma^z_i \sigma^z_{j} -
\sum_i h_i \sigma^x_i \ .
\label{ham}
\end{equation}
Here the $\{\sigma^\alpha_i\}$ are Pauli spin matrices, and the
nearest neighbor
interactions $J_{ij}$ and transverse fields $h_i$ are both independent
random variables. 
This model should provide a reasonable description of
the experimental system\cite{rosen}
LiHo$_x$Y$_{1-x}$F$_4$ and may also be
an appropriate model\cite{neto}
to describe non-fermi liquid behavior in certain $f$-electron
systems.

Naturally the random transverse field Ising model has been quite
extensively studied and many surprising {\em analytical} results are
available\cite{dsf,sm,mw} for the case of dimension $d=1$.  For
example, the dynamic critical exponent, $z$, is infinite.  Instead of
a characteristic time scale $\xi_\tau$ varying as a power of a
characteristic length scale $\xi$ according to $\xi_\tau \sim \xi^z$,
one has instead an exponential relation\cite{dsf} $ \xi_\tau \sim \exp
({\rm const.}\ \xi^\psi), $ where $\psi = 1/2$. This is called {\em
  activated} dynamical scaling.  In addition, distributions of the
equal-time $\sigma^z_i$--$\sigma^z_{i+r}$ correlations, are very
broad. As a result {\em average} and {\em typical}\/\cite{typical}
correlations behave rather differently, since the average is dominated
by a few rare (and hence {\em atypical}\/) points. At the critical
point, for example, the average correlation function falls off with a
power of the distance $r$ as $ C_{\rm av}(r) \sim r^{-\tilde{\eta}} ,
$ where\cite{dsf} $\tilde{\eta} = (3-\sqrt{5})/2 \simeq 0.38$ whereas
the typical value falls off much faster, as a stretched exponential $
C_{\rm typ}(r) \sim \exp (- {\rm const. }\ r^\sigma ) , $ with $\sigma
= 1/2$.  As the critical point is approached, the average and typical
correlation {\em lengths} both diverge but with {\em different}
exponents\cite{dsf}, i.e.  $ \xi_{\rm av} \sim \delta^{-\nu_{\rm av}};
\ \xi_{\rm typ} \sim \delta^{-\nu_{\rm typ}}, $ where $\delta$ is the
deviation from criticality, and $ \nu_{\rm av} = 2, \nu_{\rm typ} =
1.$ Finally, there are strong Griffiths-McCoy singularities at low
temperature even away from the critical point, coming from rare
regions which are ``locally in the wrong phase''.  These are
characterized by a single continuously varying
exponent\cite{dynamical}, $z^\prime(\delta)$, which {\em diverges} as
$\delta \to 0$.

An important question is whether these striking analytical results 
are a special feature
of 1-d, or whether they are valid more generally.
Unfortunately, the analytical approach is only valid in 1-d, and very little
is known in higher dimensions.
Senthil and Sachdev\cite{ss} have studied the model in Eq.~(\ref{ham})
with site dilution and shown that activated dynamical scaling occurs along
that
part of the zero temperature phase boundary which is precisely {\em at} the
percolation concentration. However, it is not clear if
this result also holds for the rest of the phase boundary, and, to our
knowledge, there are no results at all for other, more general, models.
Here, we 
investigate the behavior of the random transverse field Ising ferromagnet
in {\em two} dimensions by 
performing large-scale Monte Carlo simulations.
Because
the ferromagnet has no frustration we are able to use highly efficient cluster
algorithms
which considerably reduce critical slowing down.
Our main conclusion is that the behavior of the 2-d system
{\em is} very similar to that of 1-d.

In order to capture the random quantum critical behavior in the
intermediate size
systems that we can simulate,
we wish the disorder to be effectively quite strong. In particular, we
would like some of the fields to be {\em much} stronger than the bonds in their
vicinity and vice-versa, which is captured by having distributions
for both the fields and interactions with a finite weight at the origin.
We therefore take the following ``box'' distribution
\beqn
\pi(J_{ij})&=&\cases{1,&for $0<J_{ij}<1$\cr
                0,&otherwise\cr}\;\nonumber\\
\rho( h_i)&=&\cases{ h^{-1},&for $0< h_i< h$\cr
                0,&otherwise\cr}\;.
\label{uniform}
\eeqn

As is standard\cite{suzuki_kogut}, we represent the $d$-dimensional
quantum Hamiltonian in
Eq.~(\ref{ham}) by an effective classical action in
($d$+1)-dimensions, where the extra dimension, imaginary time, is of size
$\beta \equiv 1/T$,
and is divided up into $L_\tau \equiv \beta/\Delta\tau$
intervals each of width $\Delta\tau$ in the limit
$\Delta\tau \to 0$. The partition function can then be written
as
$Z = \lim_{\Delta\tau\to 0} {\rm Tr} \exp(-S)$,
where the effective classical action is given by
\begin{equation}
S = -\sum_{\langle i, j\rangle, \tau} K_{ij}
S_i(\tau) S_j(\tau) - \sum_{i, \tau} \widetilde{K}_i S_i(\tau) S_i(\tau^\prime)
\end{equation}
where, $\tau^\prime = \tau + \Delta \tau$, $S_i(\tau)=\pm 1$,
\begin{equation}
 K_{ij} = \Delta \tau J_{ij}, \quad {\rm and } \quad \exp(-2\widetilde{K}_i) =
\tanh(\Delta \tau h_i) . 
\label{mapping}
\end{equation}

To study large systems sizes with small statistical errors, we use
cluster algorithms which simultaneously flip many spins. For our
results on Griffiths-McCoy singularities we have developed\cite{rk} a
variant of the loop algorithm\cite{bw} in which the required limit
$\Delta\tau \to 0$ is explicitly taken. We shall call this the {\em
continuous imaginary time} algorithm. It represents the original
quantum Hamiltonian exactly (apart from statistical errors). We tune through
the transition by varying $h$. 

In our simulations which determine for critical exponents, we use a
different approach, and exploit universality according to which the
universal quantities should be independent of $\Delta\tau$ and so, for
convenience, we set $\Delta\tau = 1$. We then have a three-dimensional
Ising model, with disorder perfectly correlated in one direction,
which we simulate using the Wolff\cite{wolff} cluster algorithm. We
shall call this the {\em discrete imaginary time} algorithm. It does
{\em not} represent the quantum Hamiltonian exactly but is expected to
be in the same universality class.  Furthermore, for this algorithm we
find it convenient to parameterize the strength of fluctuations by an
effective classical temperature $T_{cl} \equiv 1/\beta_{cl}$, (not
equal to the real temperature which is the inverse of the size in the
time direction) and write $Z = {\rm Tr} \exp(-\beta_{cl} {\cal
  H}_{cl})$ where
\begin{equation}
{\cal H}_{cl} = -\sum_{\langle i, j\rangle, \tau} J_{ij}
S_i(\tau) S_j(\tau) - \sum_{i, \tau} \widetilde{J}_i S_i(\tau) S_i(\tau + 1) .
\end{equation}
Here $\tau$ runs over integer values, $1\le \tau \le L_\tau$, and the
distributions of the interactions are given by
\begin{eqnarray}
\pi(J_{ij}) & = &
\left\{
\begin{array}{ll}
1 & \mbox{for $ 0 < J_{ij} < 1$,} \\
0  & \mbox{otherwise},
\end{array}
\right.
\nonumber \\
\rho(\widetilde{J}_i) & = &
\left\{
\begin{array}{ll}
2 \exp(-2 \widetilde{J}_i)  & \mbox{for $  \widetilde{J}_i \ge 0$,} \\
0  & \mbox{for $  \widetilde{J}_i < 0$,}
\end{array}
\right.
\label{dist}
\end{eqnarray}
which, from Eq.~(\ref{mapping}), is similar to Eq.~(\ref{uniform}).
We tune through the transition by varying the classical temperature, $T_{cl}$.

The lattice is of size $L$ in the space directions, and
we denote disorder averages by $[\cdots]_{\rm av}$
and Monte Carlo averages
by $\langle \cdots
\rangle$. For both algorithms we employ periodic boundary conditions in all
directions.

\begin{figure}
\epsfxsize=\columnwidth\epsfbox{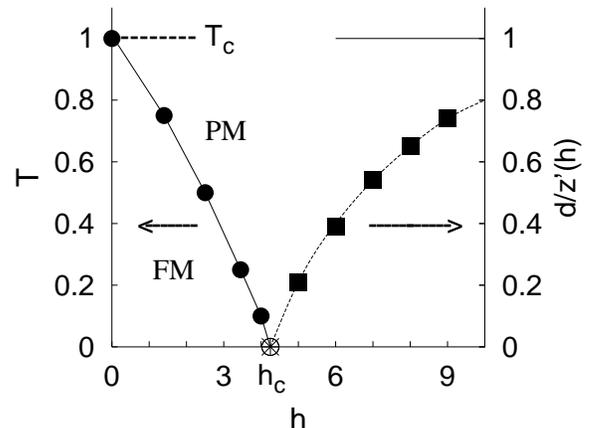}
\caption{
\label{figall}
Results obtained with the continuous imaginary time algorithm.  The
left hand axis indicates the phase diagram of the $d=2$ random
transverse Ising model: PM means paramagnetic, FM means ferromagnetic,
$T_c=1.00(1)$ is the critical temperature of the classical random
Ising ferromagnet ($h_i=0$) with the box bond distribution in
Eq.~(\protect\ref{uniform}),
and $ h_c=4.2(2)$
the location of the quantum critical point we are interested in.  The
right hand scale indicates the values of $d/z'(h)$ obtained from
analyzing the integrated probability distribution of $\ln\chi_{\rm
local}$ according to (\protect{\ref{localsus}}) in the
Griffiths-McCoy region, $h > h_c$. The open circle corresponds to
$z'(h_c)=\infty$ and the horizontal line at $d/z'=1$ indicates the
expected limit $\lim_{ h\to\infty}z'(h)=d$. The broken line is just a
guide to the eye.}
\end{figure}

First of all we discuss our results for the location of the critical line
using the continuous imaginary time algorithm.
We do this by computing the Binder ratio
\begin{equation}
g_{\rm av} = {1 \over 2} \left[ 3 - {\langle M^4\rangle \over \langle
M^2\rangle^2}
\right]_{\rm av} ,
\label{binder}
\end{equation}
where $M = \sum_{i} \int_0^\beta  S_i(\tau) \, d\tau $,
At any finite temperature the
system is expected to be in the universality class of the classical
two-dimensional classical random bond Ising ferromagnet. For small
temperatures the
size of the classical critical region shrinks and we need to study
larger sizes (we went up to $L=32$) to get a reliable estimate of $
h_c(T)$. By extrapolating the latter to $T=0$, see Fig. \ref{figall},
we obtain for the location of the quantum critical point $ h_c=
h_c(T=0)=4.2\pm0.2$.

Now we turn our attention to the Griffiths-McCoy region in the
disordered phase ($ h> h_c$). Due to the presence of
strongly coupled regions in the system the probability distribution of
excitation energies (essentially inverse tunneling times for these
ferromagnetically ordered clusters) becomes extremely broad. As a
consequence we expect the probability distribution of local
susceptibilities to have an algebraic tail at $T=0$
\cite{yr,igloirieger,qsggriff,rygbhgriff}:
\be
\Omega(\ln\chi_{\rm local})
\approx-\frac{d}{z'(h)}\ln\chi_{\rm local} ,
\label{localsus}
\ee
where $\Omega(\ln\chi_{\rm local})$ is the probability for the
logarithm of the local susceptibility $\chi_i$ at site $i$ to be
larger than $\ln\chi_{\rm local}$. The dynamical exponent\cite{dynamical}
$z'(h)$
varies continuously with the distance from the critical point and
parameterizes the strengths of the Griffiths-McCoy singularities also
present in other observables. At finite temperatures the distribution
of $\chi_{\rm local}$ is chopped off at $\beta$, and close to the critical
point one expects finite size corrections as long as $L$ or $\beta$
are smaller than the spatial correlation length or imaginary
correlation time, respectively. We used $\beta\le1000$ and averaged
over at least 512 samples.

In Fig. \ref{figall} we show our results for $d/z'(h)$ in the
Griffiths-McCoy region. For $ h\to\infty$ we expect $d/z'(h)\to1$,
since this is the result for {\it isolated} spins in random fields
with non-vanishing probability weight at $ h_i=0$. The more
interesting limit is $ h\to h_c$. The data are well compatible with
$\lim_{ h\to h_c} z'(h)=\infty$, as in one-dimension\cite{dsf,yr}. 
The average susceptibility $[\chi]_{\rm av}$
diverges like\cite{rygbhgriff} $[\chi]_{\rm av}\sim T^{d/z'(h)-1}$ for
$T\to 0$. Hence, if $\lim_{ h\to h_c} z'(h)$ is universal, {\em i.e.}\/ does
not depend on the details of the disorder,
as is the case in 1-d,
$[\chi]_{\rm av}$ diverges quite generally
in a {\em range} about the quantum critical point for systems with 
Ising symmetry.

Next we describe our results for critical exponents using the discrete
imaginary time algorithm, for which
we studied sizes up to $L=48$ and $L_\tau=2048$.
We found that no more than
100 sweeps were required for equilibration, even for the largest
lattices. At least 1000 samples were averaged over.

\begin{figure}
\epsfxsize=7cm\epsfbox{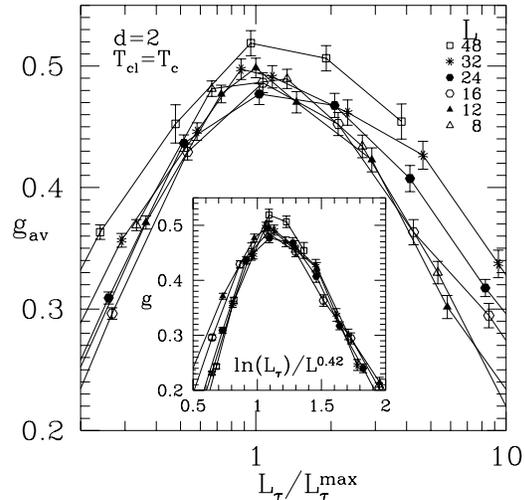}
\caption{
Results using the discrete imaginary time algorithm
at the quantum critical point, $T_{cl} = T_c^{cl} = 2.45$.
In the main
figure the horizontal axis is
$L_\tau/L_\tau^{\rm max}$ where $L_\tau^{\rm max}$ is
the value of $L_\tau$ at the peak. Note that the curves do not scale but
rather get broader for larger
sizes, indicating activated scaling, $z=\infty$,
The data for $L=48$ is slightly high which may indicate that the true value of
$T_c^{cl}$ is a little higher. In the inset, the data for $L \ge 12$ is seen to
scale quite
well with the same form $\ln L_\tau  / L^{\psi}$ known to be exact in 1-d, but
the value of $\psi=0.42$ (shown)
works better than the 1-d value of $\psi=1/2$. 
}
\label{g-2d}
\end{figure}

We locate the $T=0$ critical point by a method already used for the
quantum spin glass\cite{rygbh}. We compute the Binder ratio,
Eq.~(\ref{binder}) ,
which (assuming, for now, that $z$ is finite) has the finite-size
scaling form \begin{equation} g_{\rm av} = \widetilde{g}\left( \delta
L^{1/\nu_{\rm av}}, L_\tau/L^z \right) ,
\label{g_fss} \end{equation} where $\delta = T_{cl} - T_{cl}^c$, with
$T_{cl}^c$ the value of $T_{cl}$ at criticality.  For fixed $L$,
$g_{\rm av}$ has a peak as a function of $L_\tau$. At the critical
point, $T_c^{cl}$, the peak height is independent of $L$ and the
values of $L_\tau$ at the maximum, $L_\tau^{\rm max}$, vary as
$L^z$. Furthermore, a plot of $g_{\rm av}$ against $L_\tau/L_\tau^{\rm
max}$ at the critical point, which has the advantage of not needing a
value for $z$, should collapse the data. We see in Fig.~\ref{g-2d}
that this does {\em not} happen. Rather the curves clearly become
broader for larger sizes.  This is easy to understand since we know
that for 1-d $z$ is infinite and it is the {\em log} of the
characteristic time which scales with a power of the length scale.
This suggests that the scaling variable should be $\ln L_\tau / \ln
L_\tau^{\rm max}$ with $\ln L_\tau^{\rm max} \sim L^\psi$, say.  A
corresponding scaling plot is shown in the inset to
Fig.~\ref{g-2d}. The data collapse for sizes $L \ge 12$, quite good
for $\psi=0.42$, and not quite so good with the 1-d value,
$\psi=1/2$, though we would not claim that $\psi=1/2$ is ruled out. We
have also performed analogous calculations for one-dimension\cite{py}
with very similar results.  The close similarity of the data for 1-d
and 2-d, and the broadening of the data in the main part of
Fig.~\ref{g-2d}, suggests that $z$ is infinite in 2-d, as well as in
1-d.

\begin{figure}
\epsfxsize=7cm\epsfbox{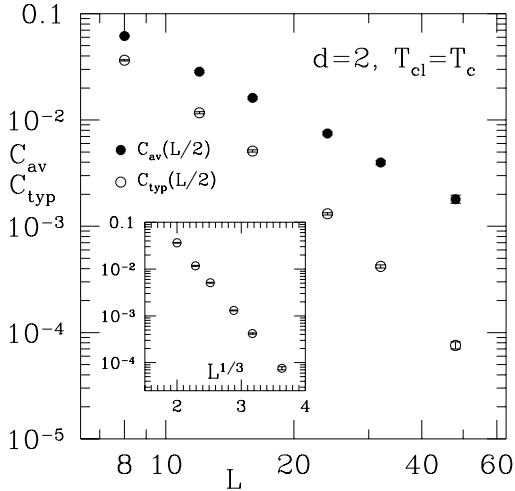}
\caption{
The main figure shows the average and typical\protect\cite{typical}
correlations
between spins $L/2$ apart at the critical point. The average falls off with a
power law,
and a fit gives a slope of $-\tilde{\eta}$ with $\tilde{\eta}=1.95$.
However,
the curvature of the data for the
typical correlation function shows that this falls off {\em faster}
than a power law. 
The inset shows that the 
data is consistent with a stretched
exponential form, $\exp (-{\rm const.}\ L^\sigma)$, but with $\sigma=1/3$
rather than the value of 1/2 found\protect\cite{dsf} in 1-d.
}
\label{cf-2d}
\end{figure}

Next we consider the equal time correlations at the
critical point.  
Fig.~\ref{cf-2d} shows data for the average and typical\cite{typical}
correlations for spins separated by $\vec{r} = (L/2, 0)$. For each
value of $L$, we took $L_\tau$ such that $g_{\rm av}$ is close to the
peak\cite{L-Ltau} shown in Fig.~\ref{g-2d}.  According to finite size
scaling, the dependence on $L$ for a finite system should be the same
as the dependence on $r$ in a bulk system.

The data in Fig.~\ref{cf-2d} shows that the average correlation
function falls off with a power law, with $\tilde{\eta}$ about $1.95$,
while the typical value falls off faster than a power law, (because of
the downward {\em curvature}) consistent with a stretched exponential
behavior of the form $\exp(-{\rm const.}\ L^\sigma)$, with $\sigma
\simeq 1/3$. The statistical errors (shown) are generally smaller than
the size of the points, so the downward curvature is statistically
significant. This behavior is of the form expected in one
dimension\cite{dsf}, except that there $\sigma=1/2$, a result which is
reproduced by our 1-d simulations\cite{py}.
Moreover, in one dimension
$\psi=\sigma$, and this relation also holds in
higher dimensions\cite{privcomm} provided that the fixed point is similar,
{\em i.e.}\/ has infinitely strong disorder.
Our data is compatible with this result 
though neither $\sigma$ nor $\psi$ are determiend with precision.
Additional results, including the whole distribution of
correlation functions, will be presented in a separate publication.

To conclude, we have found a strong similarity between the critical
behavior of the random transverse field Ising model in one and two
dimensions. In particular, both $z$ and $\lim_{h\to h_c} z^\prime(h)$
are infinite. Previous simulations on quantum spin
glasses\cite{rygbhgriff} (for which the mean of the distribution of
the $J_{ij}$ is zero) in two dimensions, found these quantities to be
apparently finite. However, it is plausible that the asymptotic result
should be infinite also for quantum spin glasses, and that the finite
result found is an artifact of the smaller sizes used there. 
(Those studies also used a non-random transverse field.)

Two of us (NK and HR) have implemented numerically for $d=2$ a natural
generalization of the renormalization group procedure used in Ref.\ 
\onlinecite{dsf} for $d=1$.  After this work was completed, we heard
that S.-C. Mau, O. Motrunich and D. A. Huse (private communication)
used a similar approach and observed a flow to the infinite disorder
critical fixed point, just as in $d=1$.

We would like to thank D.~S.~Fisher, D.~A.~Huse,
R.~N.~Bhatt and F.\ Igl\'oi for helpful discussions.
This work was supported by the National Science Foundation under grant DMR
9713977 (APY), the Deutsche Forschungsgemeinschaft (DFG) under contract Pi
337/1-2 (CP), the DFG Grant number Ri-580/4-5 (HR), and 
Grant-in-Aid for Scientific Research Program 
(No. 09740320) from the Ministry of Education, Science, Sport and Culture
of Japan (NK). The work of CP and APY was also supported by an allocation of
computer time from the Maui High Performance Computing Center.
HR also thanks Toho University Department of Physics for kind
hospitality.


\begin{references}
\bibitem{dynamical}
The exponent $z^\prime$ arises because there are
localized cluster
excitations with low energy coming from regions which are locally in the wrong
phase. These have a power law density of states such that the lowest energy in
a region of linear size $L$ (and hence volume $L^d$) is of order
$1/L^{z^\prime}$. Hence $z^\prime$ can be viewed as
a dynamical exponent since it relates an energy scale to a length scale.
\end{references}
\end{document}